# Generalized Master-Slave-Splitting Method and Application to Transmission-Distribution Coordinated Energy Management

Zhengshuo Li, *Member, IEEE*, Hongbin Sun, *Fellow, IEEE*, and Qinglai Guo, *Senior Member, IEEE*

*Abstract*—Transmission-Distribution coordinated energy management (TDCEM) is recognized as a promising solution to the challenge of high DER penetration, but there is a lack of a distributed computation method that universally and effectively works for the TDCEM. To bridge this gap, a generalized master-slave-splitting (G-MSS) method is presented in this paper. This method is based on a general-purpose transmission-distribution coordination model called G-TDCM, which thus enables the G-MSS to be applicable to most of the central functions of the TDCEM. In this G-MSS method, a basic heterogenous decomposition (HGD) algorithm is first derived from the HGD of the coupling constraints in the optimality conditions of the G-TDCM. Its optimality and convergence properties are then proved. Further, inspired by the conditions for convergence, a modified HGD algorithm that utilizes the subsystem's response function is developed and thus converges faster. The distributed G-MSS method is then demonstrated to successfully solve a series of central functions, e.g. power flow, contingency analysis, voltage stability assessment, economic dispatch and optimal power flow, of the TDCEM. The severe issues of over-voltage and erroneous assessment of the system security that are caused by DERs are thus resolved by the G-MSS method with modest computation cost.

*Index Terms*—Distributed energy resource, distributed optimization, distribution, energy management, transmission.

## I. INTRODUCTION

### A. Backgrounds

IT is well known that high penetration of distributed energy resources (DERs) challenges both distribution (DPS) and transmission system (TPS) operation. A research group at MIT and a working group of IEEE have identified that for a country with high penetration of DERs, the generation at a DPS "could impact a country's transmission system" [1] and "a closer cooperation between transmission system operators (TSOs) and distribution system operators (DSOs) is imperative" [2].

This work was supported by the China Postdoctoral Science Foundation under Grants 2016M600091 and 2017T100078.

Zhengshuo Li is with Shenzhen Environmental Science and New Energy Technology Engineering Laboratory, Tsinghua-Berkeley Shenzhen Institute (TBSI), Tsinghua University, Shenzhen, Guangdong, 518055, China (email: shuozhengli@sina.com).

Hongbin Sun and Qinglai Guo are both with TBSI and the Department of Electrical Engineering, State Key Laboratory of Power Systems, Tsinghua University, Beijing, 100084, China (e-mail: shb@tsinghua.edu.cn).

Moreover, DERs could also improve the power system operation in the congestion mitigation, voltage support, etc. [3], [4]. The transmission-distribution (T-D) coordination is also essential, and even "of utmost importance" [1], to realize these services provided by DERs. Nevertheless, as reviewed in [4], while there are some works on the T-D coordination, e.g. the network code of ENTSO-E [5], "most limit to establish very basic principles without providing concrete coordination mechanisms" [4]. Therefore, a thorough study of the T-D coordination mechanism, which will be beneficial to TSOs and DSOs, becomes necessary and is carried out in this paper.

### B. Literature Review

Most works on the T-D coordination could be classified into the following subjects:

1) Modelling and dynamic simulation of an integrated transmission and distribution (ITD) system [6]-[11]: especially, a distributed master-slave-splitting (MSS) method was proposed in [7] to solve the ITD power flow (PF) equations.

2) Coordinated voltage control at the boundary of the TPS and the DPS: passive control was first investigated to reduce the impact of DERs on the TPS [12]; then active control where TSOs and DSOs are coordinated was studied in [13]-[17], in which both heuristic methods [13]-[15] and mathematical decomposition methods [16]-[17] were experimented. Recently, [18] showed that using transformer tap stagger in a DPS improves the DPS's voltage support capability, which provides another measure that can be taken in the T-D coordination.

3) Voltage stability assessment (VSA) of the ITD system: initially, the impact of the load tap changers (LTC) on the static voltage stability of an ITD system was noticed [19]; several years later, the impact of the DERs on the stability was investigated [20]-[23], and recently a distributed method was proposed in [23] to assess the critical point of the ITD system.

4) Active power dispatch: [24] studied a distributed solution to a unit commitment problem using the so-called ATC method; [25] and [26] studied distributed solutions to an economic dispatch (ED) problem based on the heterogenous decomposition (HGD) algorithm and parametric programming, respectively; [27] investigated a coordinated market in the context of T-D coordination.

### C. Contributions

As can be seen from the above, T-D coordinated energy management (TDCEM) that enables distributed cooperation between TSOs and DSOs is a promising solution to the challenge of high DER penetration. In this regard, however, there remains an important but unsolved problem in the literature: "Is there a distributed computation method that universally and effectively works for the central functions, e.g. PF, VSA, ED, optimal power flow (OPF), etc., of a TDCEM system?" [28] Remarkably, developing such a method is not only of theoretical value but also of notable practical significance: it will demonstrate a universally effective cooperation mechanism for TSOs and DSOs, providing sound answers to the questions like what the respective energy management of TPSs and DPSs should be modified for the T-D coordination and what are the minimum data to be shared among TSOs and DSO to accomplish this coordination [28].

To bridge this gap, this paper presents a distributed generalized master-slave splitting (G-MSS) method that comes from the first author's Ph.D. dissertation [28], showing its optimality and convergence properties and demonstrating its application to the TDCEM. This research expands our previous works in the following aspects:
- Relative to the MSS method that is only applicable to the PF problem [7], the G-MSS method is applicable to a general-purpose continuous optimization model involving ED, OPF, etc.
- Relative to [17] and [25] that use an HGD algorithm, this paper presents a new algorithm called modified HGD that converges faster and has a larger domain of convergence.
- Relative to [29], we mathematically analyze for a general case why introducing a response function in the iteration accelerates the convergence, and thereby develop a generally workable algorithm.

In other words, and as will be seen in the sequel, those previous works are only special case of the G-MSS method, which will have a wider range of application.

The remainders are as follows. In Section II, a generalized transmission-distribution coordination model (G-TDCM) will be established. In Section III, two distributed algorithms in the G-MSS method are presented with proved optimality and convergence properties. In Section IV, the application of the G-MSS method to the PF calculation, contingency analysis (CA), VSA, ED and OPF is demonstrated. Finally, conclusions are presented.

## II. GENERALIZED TRANSMISSION-DISTRIBUTION COORDINATION MODEL

A G-TDCM means that it is applicable to most of the central functions of a TDCEM system. Hence, it allows one to develop a universally applicable distributed solution to the TDCEM instead of designing algorithms for every specific function, which will thus save effort in establishing a TDCEM system and help to reveal the basic coordination rules.

To commence with this model, an ITD system is divided into a master, a boundary and a slave subsystem. The boundary refers to the interface of the TPS and the DPS, typically a high-voltage or low-voltage bus of a distribution substation. The master and slave subsystems consist of the other components (e.g., buses, lines, generator, loads, etc.) of the TPS and the DPS, respectively, which are separately supervised by the TSO and the DSO. In addition, the practice in power system operation reveals the following facts:

**Fact 1**: The control in the boundary subsystem is decided by either a TSO or a DSO.

**Fact 2**: The master and slave subsystems are coupled by the state of the boundary subsystem. In other words, any power flow path that connects the master and slave subsystems must pass through the boundary subsystem, as is illustrated in Fig. 1.

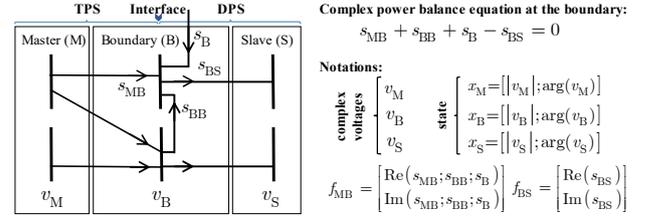

Fig. 1. The power flow inside the boundary subsystem.

As for this master-slave-structured ITD system, let $z$ be the optimal variable that contains both $x$, the state of the system, and $u$, the control, and let $f$, $g$ be the equality and inequality function, respectively. Then, the G-TDCM is formulated as a general-purpose continuous optimization problem below [28]:

$$\min_{z_M, x_B, z_S} \left\{ \begin{array}{l} c_M(z_M, x_B) \\ + c_S(x_B, z_S) \end{array} \middle| \begin{array}{l} f_M(z_M, x_B) = 0, \\ f_B(z_M, x_B, z_S) = 0, f_S(x_B, z_S) = 0, \\ g_M(z_M, x_B) \geq 0, \quad g_S(x_B, z_S) \geq 0 \end{array} \right\} \quad (1)$$

where the subscripts M, B and S denote the master, boundary and slave subsystem variables and functions, respectively. Notice that, due to Fact 1, the control in the boundary subsystem, $u_B$, is contained by either $z_M$ or $z_S$.

In (1), the objective can represent minimizing generation costs or many other common operation targets, but it seems to be limited to a family of functions that do not contain in the formula a master-slave coupling term like $c_B(z_M, x_B, z_S)$. Indeed, such a $c_B$ can be involved in (1) if it can be written as $\mu(\nu_{MB}(z_M, x_B) + \nu_{BS}(x_B, z_S))$ where $\mu$, $\nu_{MB}$ and $\nu_{BS}$ are functions[1]. In addition, the equality constraints in (1) are restricted to only power flow equations, the very ones that couple the TPS and the DPS. The possible other equality constraints, e.g. those regarding the operating of a device $\alpha$, are modeled by $g_\alpha(z) \geq 0$ and $-g_\alpha(z) \geq 0$, and thus represented by the inequality constraints in (1).

As can be seen from Fact 2 and Fig. 1, the master-slave structure of an ITD system enables one to reformulate $f_B(z_M, x_B, z_S)$ as the difference of a function $f_{MB}(z_M, x_B)$ and a function $f_{BS}(x_B, z_S)$, namely $f_B = f_{MB} - f_{BS}$ [2]. In other words, this structure helps to reveal the physical feature implicit in the formula $f_B(z_M, x_B, z_S)$, which will facilitate the following G-MSS method.

---
[1] The proof is omitted here because of space limitations.
[2] This can be proved via analyzing the complex power flow equations regarding the boundary subsystem, but the proof is omitted because of space limitations.

## III. Generalized Master-Slave-Splitting Method [28]

### A. Heterogeneous Decomposition of Optimality Conditions

Let $L$ below denote the Lagrangian of the G-TDCM:

$$L = c_M + c_S - \lambda_M^T f_M - \lambda_B^T (f_{MB} - f_{BS}) - \lambda_S^T f_S - \omega_M^T g_M - \omega_S^T g_S \quad (2)$$

where $\lambda$ and $\omega$ are the multipliers regarding the equality and inequality constraints, respectively, and $\omega \geq 0$; superscript T stands for transpose. Note that $f_B = f_{MB} - f_{BS}$ is used in (2).

With certain constraint qualifications, the Karush–Kuhn–Tucker (KKT) conditions of the G-TDCM hold and are formulated in a canonical form below:

$$\begin{cases} l_M(\xi_M, \xi_B) = 0, \; l_B(\xi_M, \xi_B, \xi_S) = 0, \; l_S(\xi_B, \xi_S) = 0, \\ f_M(z_M, x_B) = 0, \; f_B(z_M, x_B, z_S) = 0, \; f_S(x_B, z_S) = 0, \\ g_M(z_M, x_B) \geq 0, \; g_S(x_B, z_S) \geq 0, \; \omega_M \geq 0, \; \omega_S \geq 0, \\ \omega_M^T g_M(z_M, x_B) = 0, \; \omega_S^T g_S(x_B, z_S) = 0 \end{cases} \quad (3)$$

where the variables $\xi_M := [z_M; \lambda_M; \omega_B]$, $\xi_B := [x_B; \lambda_B]$ and $\xi_S := [z_S; \lambda_S; \omega_S]$ contain the primal and dual variables of every subsystem, respectively; $l_M$, $l_B$ and $l_S$ are the partial derivatives of $L$ with regard to $z_M$, $x_B$ and $z_S$, respectively.

In (3), both $f_B$ and $l_B$ couple the variables of the master and slave subsystems. Given $f_B = f_{MB} - f_{BS}$, $l_B$ can also be reformulated as $l_B(\xi_M, \xi_B, \xi_S) = l_{MB}(\xi_M, \xi_B) - l_{BS}(\xi_B, \xi_S)$, where $l_{MB} = \frac{\partial c_M}{\partial x_B} - \left(\frac{\partial f_M}{\partial x_B}\right)^T \lambda_M - \left(\frac{\partial g_M}{\partial x_B}\right)^T \omega_M - \left(\frac{\partial f_{MB}}{\partial x_B}\right)^T \lambda_B$ and $l_{BS} = -\frac{\partial c_S}{\partial x_B} - \left(\frac{\partial f_{BS}}{\partial x_B}\right)^T \lambda_B + \left(\frac{\partial f_S}{\partial x_B}\right)^T \lambda_S + \left(\frac{\partial g_S}{\partial x_B}\right)^T \omega_S$. Based on these difference-type formulae of $f_B$ and $l_B$, the KKT conditions in (3) are decomposed into two parts, i.e. the (KKT-M) in (4) and the (KKT-S) in (5).

KKT-M: $\begin{cases} h_M(\xi_M, \xi_B) = 0, \; h_{MB}(\xi_M, \xi_B) = y_B, \\ \omega_M^T g_M(z_M, x_B) = 0, \; g_M(z_M, x_B) \geq 0, \; \omega_M \geq 0 \end{cases} \quad (4)$

KKT-S: $\begin{cases} h_S(\xi_B, \xi_S) = 0, \; \omega_S^T g_S(x_B, z_S) = 0, \\ g_S(x_B, z_S) \geq 0, \; \omega_S \geq 0 \end{cases} \quad (5)$

where $h_M := [f_M; l_M]$, $h_S := [f_S; l_S]$, $h_{MB} := [f_{MB}; l_{MB}]$, and $y_B$ is the value of the function $h_{BS} := [f_{BS}; l_{BS}]$ of a given pair $(\xi_B, \xi_S)$. Because of the asymmetric formulae of the (KKT-M) and the (KKT-S), we call this decomposition heterogeneous decomposition.

Recalling the master-slave structure of an ITD system, one can see that the (KKT-M) part with a given $y_B$ is only concerned with the variables of the TPS, and that the (KKT-S) part with a given $\xi_B$ is only concerned with the variables of the DPS. Hence, if $y_B$ and $\xi_B$ are given, the primal and dual variables of the TPS and the DPS can be independently solved from the (KKT-M) and the (KKT-S), respectively. Furthermore, these solved variables satisfy the KKT conditions in (3) if $y_B$ and $\xi_B$ are consistent, i.e., $y_B$ equals the counterpart that is produced by (5) with this $\xi_B$, and $\xi_B$ equals the counterpart that is produced by (4) with this $y_B$[3]. To arrive at this "consistency" point, two iterative algorithms are designed below. We call these algorithms the basic and modified HGD algorithms, respectively, because they are based on the above heterogenous decomposition of the KKT conditions.

### B. Basic HGD Algorithm

#### 1) Computation procedures

The basic HGD algorithm is an iterative algorithm. In every iteration, a TSO and a DSO should solve a transmission subproblem (T-SP) formulated in (6) and a distribution subproblem (D-SP) in (7), respectively:

$$\min_{z_M, x_B} c_M(z_M, x_B) - (l_{BS}^{sp})^T x_B$$
$$s.t. \begin{cases} f_M(z_M, x_B) = 0, \; g_M(z_M, x_B) \geq 0 \\ f_{MB}(z_M, x_B) = f_{BS}^{sp}, \qquad \lambda_{MB} \end{cases} \quad (6)$$

$$\min_{z_S} c_S(x_B^{sp}, z_S) + (\lambda_{MB}^{sp})^T f_{BS}(x_B^{sp}, z_S)$$
$$s.t. \; f_S(x_B^{sp}, z_S) = 0, \; g_S(x_B^{sp}, z_S) \geq 0 \quad (7)$$

where the superscript sp denotes the specified variables in each iteration, and $\lambda_{MB}$ is the multiplier with regard to the equality constraint $f_{MB}(z_M, x_B) = f_{BS}^{sp}$.

Following the above subproblems, the procedures of this basic HGD algorithm are presented below:

| HGD Algorithm Procedures (Starting from the D-SP) | |
|---|---|
| Step 1 | a) Set the maximum iteration number $K$ and the tolerance $\varepsilon$. <br> b) Initialize $\xi_B^{sp}$ as $\xi_{B,0}^{sp} = [x_{B,0}^{sp}; \lambda_{MB,0}^{sp}]$. <br> c) Let the iteration counter $k = 1$. |
| Step 2 | a) For iteration $k$, the DSO solves (7) with a given $\xi_{B,k-1}^{sp}$, and the solution is denoted by $\xi_{S,k} = [z_{S,k}; \lambda_{S,k}; \omega_{S,k}]$. <br> b) The DSO then computes $y_{B,k}^{sp} = [f_{BS,k}^{sp}; l_{BS,k}^{sp}]$ in which $f_{BS,k}^{sp} = f_{BS}(x_{B,k-1}^{sp}, z_{S,k})$ and $l_{BS,k}^{sp} = l_{BS}(\xi_{B,k-1}^{sp}, \xi_{S,k})$. |
| Step 3 | The TSO solves (6) with a given $y_{B,k}^{sp}$ and obtains the primal and dual variables $\xi_{B,k}^{sp} = [x_{B,k}^{sp}; \lambda_{MB,k}^{sp}]$. |
| Step 4 | If $\|\xi_{B,k}^{sp} - \xi_{B,k-1}^{sp}\| < \varepsilon$, the HGD algorithm is deemed to converge. Otherwise, return to Step 2 and let $k = k + 1$ unless $k = K$. |

This HGD algorithm can also start from the T-SP and the procedures are similar to the above and thus omitted. The initialization step for online operation can be conducted via online measurements and/or the latest forecast at hand. As will be shown in Section IV, this basic HGD algorithm was applied to the T-D coordinated ED and OPF problems in [17] and [25].

---

[3] The proof of this assertion is straightforward and thus omitted to save space.

*2) Optimality and convergence*

As for the optimality, recalling that $y_B = h_{BS} = [f_{BS}; l_{BS}]$, one can prove via direct comparison that the KKT conditions of the subproblems in (6) and (7) are exactly those in (4) and (5). If this HGD algorithm converges, which implies that $y_B^{sp}$ and $\xi_B^{sp}$ are consistent, then this convergent solution satisfies the KKT conditions in (3). Therefore, it is a candidate local optimizer of the G-TDCM, and it is a local optimizer if the second-order sufficient optimality conditions are satisfied or if the G-TDCM is convex. In the latter case, this convergent solution is indeed a global optimizer.

As for the convergence, we will show that the basic HGD algorithm linearly converges in the neighborhood of a local optimizer of (1) that is denoted by $\left(\xi_{B,*}^{sp}, y_{B,*}^{sp}\right)$. Based on the sensitivity theory in optimization, it follows that in the neighborhood of $\xi_{B,*}^{sp}$ there exists a continuously differentiable function $\tilde{h}_{BS}$ such that $y_B^{sp} = \tilde{h}_{BS}\left(\xi_B^{sp}\right)$, if the local primal and dual solutions to the D-SP with $\xi_{B,*}^{sp}$ satisfy (i) the second-order sufficient optimality conditions; (ii) the strict complementarity slackness condition; and (iii) that the primal solution is a regular point. Similarly, if the local primal and dual solutions to the T-SP with $y_{B,*}^{sp}$ satisfy the similar conditions, in the neighborhood of $y_{B,*}^{sp}$ there exists a continuously differentiable function $\tilde{h}_{MB}^{-1}$ such that $\xi_B^{sp} = \tilde{h}_{MB}^{-1}\left(y_B^{sp}\right)$. Thus, $\xi_{B,*}^{sp} = \tilde{h}_{MB}^{-1}\left(\tilde{h}_{BS}\left(\xi_{B,*}^{sp}\right)\right)$, which indicates that $\xi_{B,*}^{sp}$ is the fixed point of this composite function and that the fixed-point theorem can be used to analyze the local convergence property. The above analysis is formally stated below:

**Lemma 1** [28]: *Suppose the conditions (i)-(iii) required by the sensitivity theorem hold for $\left(\xi_{B,*}^{sp}, y_{B,*}^{sp}\right)$. Also, assume that there are domains $D_M$, $D_B$ and $D_S$ such that for any $\xi_B \in D_B$, the D-SP and the T-SP have unique solutions $\xi_S = H_S(\xi_B) \in D_S$ and $\xi_M = H_M(\xi_B) \in D_M$, respectively. Define composite mappings $\tilde{h}_{BS}(\cdot) = h_{BS}(\cdot, H_S(\cdot))$ and $\tilde{h}_{MB}(\cdot) = h_{MB}(H_M(\cdot), \cdot)$ in the domain $D_B$. Then, the convergence of the basic HGD algorithm is equivalent to that the mapping $\tilde{\Phi}: D_B \subset \mathbb{R}^{n_{\xi_B}} \to D_B$ defined in (8) converges to its fixed point, if $\tilde{h}_{MB}$ has an inverse mapping in the domain $D_B$ and $\tilde{h}_{BS}(D_B) \subset \tilde{h}_{MB}(D_B)$.*

$$\tilde{\Phi} = \tilde{h}_{MB}^{-1} \circ \tilde{h}_{BS} \quad (8)$$

where the notation $\circ$ represents the composite of mappings.

Lemma 1 allows one to derive the conditions guaranteeing the local convergence of the basic HGD algorithm from the fixed-point theorem, which is formally stated below:

**Convergence Theorem** [28]: *Suppose Lemma 1 holds. The basic HGD algorithm converges linearly in the neighborhood of $\xi_{B,*}^{sp}$, if either of the conditions in (9) and (10) is satisfied:*

$$\rho\left(\left(\frac{\partial \tilde{h}_{MB}}{\partial \xi_B}\right)^{-1} \frac{\partial \tilde{h}_{BS}}{\partial \xi_B}\right) < 1 \quad (9) \qquad \left\|\left(\frac{\partial \tilde{h}_{MB}}{\partial \xi_B}\right)^{-1}\right\| \left\|\frac{\partial \tilde{h}_{BS}}{\partial \xi_B}\right\| < 1 \quad (10)$$

where $\rho(\cdot)$ stands for the spectral radius of a matrix.

This theorem can be straightforward proved via the fixed-point theorem, and the conditions in (9) and (10) can also be verified thereby. Note that the Theorem 1 in [7] is a special case of this convergence theorem.

*C. Modified HGD Algorithm*

*1) Basic idea*

Recalling $h_{MB} = [f_{MB}; l_{MB}]$, $h_{BS} = [f_{BS}; l_{BS}]$ and the difference-type formulae of $f_B$ and $l_B$, one can see that the basic HGD algorithm essentially decomposes $h_B := [f_B; l_B]$ into a difference-type $h_B = h_{MB} - h_{BS}$ from which two subproblems in (6) and (7) are constructed. If $h_B$ is decomposed in an alternative way, e.g. $h_B = h'_{MB} - h'_{BS}$, with the property

$$\left\|\left(\frac{\partial \tilde{h}'_{MB}}{\partial \xi_B}\right)^{-1}\right\| \left\|\frac{\partial \tilde{h}'_{BS}}{\partial \xi_B}\right\| \leq \beta \left\|\left(\frac{\partial \tilde{h}_{MB}}{\partial \xi_B}\right)^{-1}\right\| \left\|\frac{\partial \tilde{h}_{BS}}{\partial \xi_B}\right\| \quad (11)$$

where $0 < \beta < 1$ and $\tilde{h}'_{MB}$, $\tilde{h}'_{BS}$ are derived from $h'_{MB}$, $h'_{BS}$ as $\tilde{h}_{MB}$, $\tilde{h}_{BS}$ from $h_{MB}$, $h_{BS}$, then this new decomposition leads to a modified HGD algorithm that converges faster. Moreover, this modified HGD algorithm may also have a larger domain of convergence, because the points with $\left\|\left(\frac{\partial \tilde{h}_{MB}}{\partial \xi_B}\right)^{-1}\right\| \left\|\frac{\partial \tilde{h}_{BS}}{\partial \xi_B}\right\| \in \left(1, \frac{1}{\beta}\right)$, which are outside the domain of convergence associated with the basic HGD algorithm, enter the domain of convergence associated with $\tilde{h}'_{MB}$ and $\tilde{h}'_{BS}$.

Following the above idea, we will show a typical way of constructing $h'_{MB}$ and $h'_{BS}$, which exploits the response function of the D-SP (or T-SP) with regard to $\xi_B^{sp}$ (or $y_B^{sp}$).

*2) Construction of the new decomposition*

Suppose there exist mappings $\tilde{h}_{MB}$ and $\tilde{h}_{BS}$, as are defined in Lemma 1, associated with $h_{MB}$ and $h_{BS}$ such that $h_B = h_{MB} - h_{BS}$. Let $h'_{MB}(\xi_M, \xi_B) := h_{MB}(\xi_M, \xi_B) - a(\xi_B)$ and $h'_{BS}(\xi_B, \xi_S) := h_{BS}(\xi_B, \xi_S) - a(\xi_B)$, where $a(\xi_B)$ is a continuously differentiable function. Thus, $h_B = h'_{MB} - h'_{BS}$ holds. Further, define $\tilde{h}'_{\text{MB}} = \tilde{h}_{\text{MB}} - a$, $\tilde{h}'_{\text{BS}} = \tilde{h}_{\text{BS}} - a$, and $\tilde{\Phi}' = \tilde{h}'_{\text{BM}}{}^{-1} \circ \tilde{h}'_{\text{BS}} = (\tilde{h}_{\text{BM}} - a)^{-1} \circ (\tilde{h}_{\text{BS}} - a)$. Then we have

$$\left\|\left(\frac{\partial \tilde{h}'_{\text{MB}}}{\partial \xi_{\text{B}}}\right)^{-1}\right\| \left\|\frac{\partial \tilde{h}'_{\text{BS}}}{\partial \xi_{\text{B}}}\right\| = \left\|\left(\frac{\partial \tilde{h}_{\text{MB}}}{\partial \xi_{\text{B}}} - \frac{\partial a}{\partial \xi_{\text{B}}}\right)^{-1}\right\| \left\|\frac{\partial \tilde{h}_{\text{BS}}}{\partial \xi_{\text{B}}} - \frac{\partial a}{\partial \xi_{\text{B}}}\right\|. \quad (12)$$

It follows from (12) that if $\left\|\left(\frac{\partial \tilde{h}_{\text{MB}}}{\partial \xi_{\text{B}}} - \frac{\partial a}{\partial \xi_{\text{B}}}\right)^{-1}\right\|$ is bounded above and if $\left\|\frac{\partial \tilde{h}_{\text{BS}}}{\partial \xi_{\text{B}}} - \frac{\partial a}{\partial \xi_{\text{B}}}\right\|$ is bounded above by a small number (ideally zero), then $\left\|\left(\frac{\partial \tilde{h}'_{\text{MB}}}{\partial \xi_{\text{B}}}\right)^{-1}\right\| \left\|\frac{\partial \tilde{h}'_{\text{BS}}}{\partial \xi_{\text{B}}}\right\|$ is likely to satisfy the property in (11), which ensures that faster convergence is achieved via this new decomposition. Therefore, to accelerate the convergence, $\frac{\partial a}{\partial \xi_{\text{B}}}$ should be close to $\frac{\partial \tilde{h}_{\text{BS}}}{\partial \xi_{\text{B}}}$ that is the sensitivity, or "response", of the output of the D-SP with regard to the input $\xi_{\text{B}}^{\text{sp}}$. Thus, $\frac{\partial a}{\partial \xi_{\text{B}}}$ is called a distribution-response function.

The above observation yields a way of constructing $\frac{\partial a}{\partial \xi_{\text{B}}}$, or equivalently $a(\xi_{\text{B}})$: let $\frac{\partial a}{\partial \xi_{\text{B}}}$ be equal to the total derivative of $h_{\text{BS}}$ with regard to $\xi_{\text{B}}$, namely $\frac{\partial h_{\text{BS}}}{\partial \xi_{\text{B}}} + \frac{\partial h_{\text{BS}}}{\partial \xi_{\text{S}}} \frac{\partial \xi_{\text{S}}}{\partial \xi_{\text{B}}}$, which can be obtained by solving the sensitivity equations derived from the (KKT-S). Furthermore, in view of the structure of $h_{\text{BS}}$, $a(\xi_{\text{B}})$ is further structured as $[a_f(x_{\text{B}}); a_l(\xi_{\text{B}})]$, where $a_f$ and $a_l$ correspond to $f_{\text{BS}}$ and $l_{\text{BS}}$, respectively (notice that $f_{\text{BS}}$ is a function of $x_{\text{B}}$). Since $f_{\text{BS}}$ is the power from a TPS to a DPS (cf. Fig. 1), $a_f$ can be physically understood as an equivalent of the (negative) DPS power injection at the boundary bus.

Based on the $h'_{\text{MB}}$ and $h'_{\text{BS}}$ constructed above, one can then construct new transmission and distribution subproblems and obtain the following computation procedures.

*3) Computation procedures*

The computation procedures of this modified HD algorithm are similar to those of the basic HGD algorithm except for Steps 2.b and 3:

- Step 2.b: In every iteration $k$, after solving the D-SP in (7), the DSO computes $y'^{\text{sp}}_{\text{B},k} = h'^{\text{sp}}_{\text{BS},k} = \left[f'^{\text{sp}}_{\text{BS},k}; l'^{\text{sp}}_{\text{BS},k}\right]$ and sends $y'^{\text{sp}}_{\text{B},k}$ to the TSO, where

$$f'^{\text{sp}}_{\text{BS},k} = f_{\text{BS}}\left(x^{\text{sp}}_{\text{B},k-1}, z_{\text{S},k}\right) - a_f\left(\xi^{\text{sp}}_{\text{B},k-1}\right), \quad (13)$$

$$l'^{\text{sp}}_{\text{BS},k} = l_{\text{BS}}\left(\xi^{\text{sp}}_{\text{B},k-1}, \xi_{\text{S},k}\right) - a_l\left(\xi^{\text{sp}}_{\text{B},k-1}\right). \quad (14)$$

- Step 3: after receiving $y'^{\text{sp}}_{\text{B},k}$, the TSO solves a new T-SP in (15) and sends $\xi^{\text{sp}}_{\text{B},k} = \left[x^{\text{sp}}_{\text{B},k}; \lambda^{\text{sp}}_{\text{MB},k}\right]$ to the DSO.

$$\min_{z_{\text{M}}, x_{\text{B}}} c_{\text{M}}(z_{\text{M}}, x_{\text{B}}) - \left(l'^{\text{sp}}_{\text{BS},k}\right)^{\text{T}} x_{\text{B}} - \left(\lambda^{\text{sp}}_{\text{MB},k-1}\right)^{\text{T}} a_f(x_{\text{B}}) - A_l(x_{\text{B}})$$
$$s.t. \quad \begin{cases} f_{\text{M}}(z_{\text{M}}, x_{\text{B}}) = 0, \quad g_{\text{M}}(z_{\text{M}}, x_{\text{B}}) \geq 0 \\ f_{\text{MB}}(z_{\text{M}}, x_{\text{B}}) - a_f(x_{\text{B}}) = f'^{\text{sp}}_{\text{BS},k}, \quad \lambda'_{\text{MB}} \end{cases}$$
(15)

where $A_l(x_{\text{B}}) = \int a_l(x_{\text{B}}, \lambda^{\text{sp}}_{\text{B},k-1}) dx_{\text{B}}$.

The convergent solution of this modified HGD algorithm must be a candidate local optimizer of the G-TDCM, which can be proved via direct comparison between the KKT conditions of (7), (15) at the convergent point and those listed in (3). The convergence of this algorithm will be better than the basic HGD algorithm if the property in (11) holds.

Alternatively, one can introduce a transmission-response function representing the response of the output of the T-SP in (6) with regard to the parameter $y^{\text{sp}}_{\text{B}}$, and construct another version of the modified HGD algorithm. This construction is in general similar to the above except that in every iteration $k$, the TSO solves the T-SP in (6) and the DSO solves a new D-SP in (16) with a given $\xi^{\text{sp}}_{\text{B},k-1}$. The optimality and convergence properties of this algorithm are same as those of the above algorithm with a distribution-response function.

$$\min_{z_{\text{S}}} \left\{ c_{\text{S}}\left(x^{\text{sp}}_{\text{B},k-1}, z_{\text{S}}\right) + \left(\lambda^{\text{sp}}_{\text{B},k-1} - b_\lambda\left(f^{\text{sp}}_{\text{BS},k-1}\right)\right)^{\text{T}} f_{\text{BS}}\left(x^{\text{sp}}_{\text{B},k-1}, z_{\text{S}}\right) \right.$$
$$\left. + B\left(x^{\text{sp}}_{\text{B},k-1}, z_{\text{S}}\right) \middle| f_{\text{S}}\left(x^{\text{sp}}_{\text{B},k-1}, z_{\text{S}}\right) = 0, \, g_{\text{S}}\left(x^{\text{sp}}_{\text{B},k-1}, z_{\text{S}}\right) \geq 0 \right\}$$
(16)

where $B(x_{\text{B}}, z_{\text{S}})$ and $b(y_{\text{B}})$ are differentiable functions such that $\frac{\partial b}{\partial y_{\text{B}}}$ is the transmission-response function and $\frac{\partial B}{\partial x_{\text{B}}} = \left(\frac{\partial f_{\text{BS}}}{\partial x_{\text{B}}}\right)^{\text{T}} b_\lambda$, $\frac{\partial B}{\partial z_{\text{S}}} = \left(\frac{\partial f_{\text{BS}}}{\partial z_{\text{S}}}\right)^{\text{T}} b_\lambda$.

*D. Discussions*

First, the G-MSS is universally applicable to the central functions of a TDCEM system, because it is designed for the G-TDCM that is a general-purpose coordination model. Moreover, it points out a universal coordination mechanism to realize this distributed T-D coordinated energy management.

Second, although what we presented above is for a one-TSO-to-one-DSO case, the G-MSS is indeed applicable to a one-TSO-to-multiple-DSO case as long as the DSOs are not directly connected with each other, which is typically the case in practice. To see this, just let the boundary subsystem involves all the boundary buses in an ITD system, and then the above derivation and assertions still hold.

Third, when TPSs and DPSs have to be separately modeled as single- and three-phase models, the G-MSS can be applied in the way that the single-phase T-SP and the three-phase D-SP are solved by the TSO and the DSO, respectively, and then the obtained single-phase $\xi_{\text{B}}$ or three-phase $y_{\text{B}}$ is converted to its three- or single-phase counterpart to be used in the next iteration. This conversation is based on the following assump-tions that usually hold in practice: (i) the three-phase boundary voltages are nearly symmetric; and (ii) a change in the boundary power injection of any phase identically affects the solution to the single-phase T-SP. The details in this conversion are however omitted due to the space limitation.

Lastly, in case that the T-SP or the D-SP occasionally becomes infeasible, slack variables accompanied by a penalty can be introduced into the subproblems to ensure the algorithm to proceed smoothly.

## IV. APPLICATION TO TDCEM

In an energy management system, the central functions regarding steady power system operation generally include PF calculation and the related CA and static VSA, state estimation (SE), ED and OPF. Distributed solutions to these functions of a TDCEM system can be designed using the G-MSS method, because the models of these functions are all special instances of the G-TDCM. Below we will present the application of the G-MSS to the functions of PF and the related CA and VSA, OPF and ED. The application to SE is briefly shown in Appendix A.

### A. PF & CA & VSA

The T-D PF (TDPF) model is a special case of (1) with a zero-valued objective and only equality constraints involved. Moreover, only the state of the system needs to be solved from the equality constraints. Thus, in every iteration, the subproblems of the basic HGD algorithm shown in (6) and (7) turn out to be (17) and (18), and $f_{\text{BS}}^{\text{sp}}$ and $x_{\text{B}}^{\text{sp}}$ are updated and exchanged between the TSO and the DSO until the change in $x_{\text{B}}^{\text{sp}}$ is smaller than $\varepsilon$ [28], which coincides with the MSS method in [7] and means that it is a special case of the G-MSS method.

$$\begin{cases} f_{\text{M}}(x_{\text{M}}, x_{\text{B}}) = 0 \\ f_{\text{MB}}(x_{\text{M}}, x_{\text{B}}) = f_{\text{BS}}^{\text{sp}} \end{cases} \quad (17) \qquad f_{\text{S}}(x_{\text{B}}^{\text{sp}}, x_{\text{S}}) = 0 \quad (18)$$

Although this basic HGD algorithm is simple and intuitive, the modified HGD algorithm using a distribution-response function is typically preferable for it converges faster. Notice that only $a_f(x_{\text{B}})$ needs to be constructed in this case. Apart from using the $\frac{\partial h_{\text{BS}}}{\partial \xi_{\text{B}}} + \frac{\partial h_{\text{BS}}}{\partial \xi_{\text{S}}} \frac{\partial \xi_{\text{S}}}{\partial \xi_{\text{B}}}$, one can also construct $a_f$ via a static network equivalencing approach below [28], which will be more convenient in field operation. Its accuracy is demonstrated in Appendix B.

Provided that the equivalent admittance of a DPS, denoted by $Y_{\text{S,eq}}$, is given, then the complex power $s_{\text{BS}}$ in Fig. 1 is formulated as follows:

$$s_{\text{BS}} = \text{diag}\{v_{\text{B}}\} \overline{Y}_{\text{B,S}} (\overline{Y}_{\text{S,S}})^{-1} \text{diag}\{v_{\text{S}}\}^{-1} \overline{s}_{\text{S}} \\ + \text{diag}\{v_{\text{B}}\} \overline{Y}_{\text{S,eq}} \overline{v}_{\text{B}} \quad (19)$$

In (19), in addition to the notations in Fig. 1, $Y_{\phi,\varphi}$ is the admittance matrix regarding the buses in the subsystems $\phi$ and $\varphi$ ( = M, B or S); $v_{\text{S}}$ and $s_{\text{S}}$ are respectively the complex voltages and power injections regarding the slave-subsystem buses; ■ stands for conjugate. Informally, in a distribution power flow solution, the magnitude of $v_{\text{S}}$ typically increases with that of $v_{\text{B}}$, so it can be expected that the term in the second line of (19) dominates the response of $s_{\text{BS}}$ with regard to $v_{\text{B}}$, namely the response of $f_{\text{BS}} = [\text{Re}(s_{\text{BS}}); \text{Im}(s_{\text{BS}})]$ with regard to $x_{\text{B}} = [|v_{\text{B}}|; \arg(v_{\text{B}})]$. Thus, $a_f$ is constructed as follows:

$$a_f := \left[ \text{Re}\left( \text{diag}\{v_{\text{B}}\} \overline{Y}_{\text{S,eq}} \overline{v}_{\text{B}} \right); \text{Im}\left( \text{diag}\{v_{\text{B}}\} \overline{Y}_{\text{S,eq}} \overline{v}_{\text{B}} \right) \right] \quad (20)$$

A distributed TDPF algorithm based on the above modified HGD algorithm is then outlined below:

| *Distributed TDPF algorithm based on the modified HGD algorithm* | |
|---|---|
| Step 1 | a) Add the distribution network equivalent $Y_{\text{S,eq}}$ to the admittance matrix of the TPS to establish $f_{\text{MB}} - a_f$ in (15). |
| | b) Initialize $x_{\text{B},0}^{\text{sp}}$. |
| Step 2 | a) For iteration $k$, the DSO solves (18) with a given $x_{\text{B},k-1}^{\text{sp}}$, and then computes $f_{\text{BS},k}^{\text{sp}}(x_{\text{B},k-1}^{\text{sp}}, x_{\text{S},k})$ and $a_{f,k} = a_f(x_{\text{B},k-1}^{\text{sp}})$ via (20). |
| | b) The DSO computes $f_{\text{BS},k}^{\prime\text{sp}} = f_{\text{BS},k}^{\text{sp}} - a_{f,k}$ to be sent to the TSO. |
| Step 3 | The TSO solves $\{f_{\text{M}}(x_{\text{M}}, x_{\text{B}})=0, f_{\text{MB}}(z_{\text{M}}, x_{\text{B}}) - a_f(x_{\text{B}}) = f_{\text{BS},k}^{\prime\text{sp}}\}$ and obtains $x_{\text{B},k}^{\text{sp}}$. |
| Step 4 | If $\|x_{\text{B},k}^{\text{sp}} - x_{\text{B},k-1}^{\text{sp}}\| < \varepsilon$, this algorithm is deemed to converge. Otherwise, return to Step 2 and let $k = k + 1$ unless $k = K$. |

The TDPF algorithm was applied to CA [30]. For a test system called 30Dl, it was found that (i) the TDPF successfully detects one dangerous contingency that is missed by the conventional CA; (ii) the TDPF avoids one false alarm that is yielded by the conventional CA; and (iii) the post-transmission-contingency security of DPSs is successfully checked by the TDPF. The improved accuracy is because the post-contingency state of the ITD system is evaluated as a whole in the TDPF. Besides, owing to the fast convergence property of the above distributed algorithm, the average number of the iterations between a TSO and a DSO for the TDPF is only 3.8. Furthermore, due to the notable difference in the parameters of TPSs and DPSs, direct application of the Newton-Raphson method sometimes fails to solve a centralized TDPF model [7], so the above distributed algorithm is preferable to solving the TDPF problem.

Similarly, the static voltage stability of an ITD system can also be more accurately assessed by a distributed continuation TDPF model where the above distributed TDPF algorithm is embedded [23]. An interesting finding is that in the context of high DER penetration, the true critical point and the loading margin of the ITD system are notably different from what the conventional transmission or distribution VSA method computes, and only the T-D VSA that concerns the whole ITD system accurately compute them in this case.

## B. OPF

A continuous T-D OPF (TDOPF) problem can be directly solved by the basic HGD algorithm given in Section III.B. Generally speaking, in every iteration, $x_{\rm B}^{\rm sp}$ and $f_{\rm BS}^{\rm sp}$, namely the boundary voltage and the power injection from the TPS to the DPS, are exchanged between the TSO and the DSO to drive the solutions to the T-SP and the D-SP to satisfy the boundary power flow equations; while $\lambda_{\rm MB}^{\rm sp}$ and $l_{\rm BS}^{\rm sp}$, representing the response of the optimum of the T-SP and the D-SP with regard to the change in the boundary-bus power and voltages, respectively, are exchanged to lead the solutions to the subproblems to a candidate local optimizer of the centralized TDOPF model. Thus, this intuitive iterative scheme guarantees the feasibility and often local optimality of an ITD system operation. More details were reported in [17].

In comparison with the uncoordinated mode where transmission and distribution OPF are separately performed, the tests on different scales of ITD systems with high DER penetration in [17] confirmed that (i) the TDOPF alleviates the distribution over-voltage issue with fewer DER curtailments; and (ii) the TDOPF mitigates the transmission congestion with lower redispatch costs by controlling DERs. This is because when there is a change in the state of a certain system, e.g. the over-voltage issue taking place in a DPS, the boundary variables, e.g. voltages or power injection, will be coordinated by the operators of the connected systems in an appropriate and often optimal way as indicated by the G-MSS method. Furthermore, the tests in [17] also indicate that the basic HGD algorithm enjoys less than about 50% of the iterations on average relative to the regular distributed optimization algorithms like APP and OCD, so the G-MSS method is computationally cheaper.

## C. ED

Like the OPF case, a T-D coordinated ED (TDCED) problem set up via DC power flow equations, whose state and control are respectively phase angles and active power, naturally exemplifies the G-TDCM. However, an ED model can also be written as a shift factor version, and this version saves the phase angles of every bus and thus typically computationally cheaper. We will show below that the TDCED problem set up via shift factor is also a special case of the G-TDCM [28], so the G-MSS method is applicable and the resultant distributed algorithms are thus preferable.

Let $P_{\rm T}$ denote the power injection in the TPS buses, $P_{\rm D}$ the injection in the DPS buses, and $P_{\rm B}$ the power flowing from the TPS to the DPS, which is "load" to the TSO and "generation" to the DSO. Then a TDCED model minimizing the generation cost of the ITD system is formulated in (21):

$$\min_{P_{\rm T},P_{\rm B},P_{\rm D}} c_{\rm T}(P_{\rm T}) + c_{\rm D}(P_{\rm D})$$
$$s.t. \begin{cases} E_{P_{\rm T}} P_{\rm T} + E_{{\rm PB}_{\rm T}} P_{\rm B} = d_{\rm T}, F_{P_{\rm T}} P_{\rm T} + F_{{\rm PB}_{\rm T}} P_{\rm B} \geq e_{\rm T} \\ E_{P_{\rm D}} P_{\rm D} + E_{{\rm PB}_{\rm D}} P_{\rm B} = d_{\rm D}, F_{P_{\rm D}} P_{\rm D} + F_{{\rm PB}_{\rm D}} P_{\rm B} \geq e_{\rm D} \end{cases} \quad (21)$$

where $E$ and $F$ denote the coefficient matrices and their subscripts denote the variables they are associated with; $d$ and $e$ denote the right-hand vectors, and the subscripts T and D denote the systems the $d$ and $e$ are associated with.

By introducing into (21) two auxiliary variables $P_{\rm BT}$ and $P_{\rm BD}$ such that $P_{\rm BT}-P_{\rm BD}=0$ to replace $P_{\rm B}$, and by defining $z_{\rm M}=[P_{\rm T},P_{\rm BT}]$, $z_{\rm S}=[P_{\rm D},P_{\rm BD}]$ and a dummy boundary state $x_{\rm B}=0$, one can convert (21) into the equivalent in (22) that is a special case of the G-TDCM:

$$\min_{z_{\rm M},x_{\rm B},z_{\rm S}} c_{\rm T}(z_{\rm M}) + c_{\rm D}(z_{\rm S})$$
$$s.t. \begin{cases} f_{\rm M}(z_{\rm M},x_{\rm B}) = [E_{P_{\rm T}} P_{\rm T} + E_{{\rm PB}_{\rm T}} P_{\rm BT} - d_{\rm T}; x_{\rm B}] = 0 \\ f_{\rm MB}(z_{\rm M},x_{\rm B}) - f_{\rm BS}(z_{\rm S}) = (P_{\rm BT} + x_{\rm B}) - P_{\rm BD} = 0 \\ f_{\rm S}(z_{\rm S}) = E_{P_{\rm D}} P_{\rm D} + E_{{\rm PB}_{\rm D}} P_{\rm BD} - d_{\rm D} = 0 \\ g_{\rm M}(z_{\rm M}) = F_{P_{\rm T}} P_{\rm T} + F_{{\rm PB}_{\rm T}} P_{\rm BT} - e_{\rm T} \geq 0 \\ g_{\rm S}(z_{\rm S}) = F_{P_{\rm D}} P_{\rm D} + F_{{\rm PB}_{\rm D}} P_{\rm BD} - e_{\rm D} \geq 0 \end{cases} \quad (22)$$

With the G-MSS method being applied to (22), it follows that $l_{\rm BS}=0$. Thus, in the basic HGD algorithm, the T-SP in (6) turns out to be an ordinary transmission ED problem with a given $P_{\rm BD}^{\rm sp}$ that is iteratively updated by the D-SP in (7) that indeed becomes (23) in this case:

$$\min_{P_{\rm D},P_{\rm BD}} c_{\rm D}(P_{\rm D}) + (\lambda_{\rm MB}^{\rm sp})^{\rm T} P_{\rm BD}$$
$$s.t. \; E_{P_{\rm D}} P_{\rm D} + E_{{\rm PB}_{\rm D}} P_{\rm BD} = d_{\rm D}, F_{P_{\rm D}} P_{\rm D} + F_{{\rm PB}_{\rm D}} P_{\rm BD} \geq e_{\rm D} \quad (23)$$

where $\lambda_{\rm MB}^{\rm sp}$ iteratively updated by the T-SP is the locational marginal price (LMP) at the boundary bus [cf. (6)] in this case. Thus, with regard to this shift-factor based model (21), the basic HGD algorithm only requires to iteratively update and exchange $\lambda_{\rm MB}^{\rm sp}$ and $P_{\rm BD}^{\rm sp}$, which, in comparison with the case of using DC power flow, saves half of the exchange data.

Moreover, this modelling also enables a simpler formulation of the preceding modified HGD algorithm. In this case, the transmission-response function $\frac{\partial b}{\partial P_{\rm BD}}$ should be equal to $\frac{\partial \xi_{\rm MB}}{\partial P_{\rm BD}}$. Since $\frac{\partial f_{\rm BS}}{\partial x_{\rm B}}=0$, $\frac{\partial B}{\partial x_{\rm B}}=0$ and $\frac{\partial B}{\partial z_{\rm S}}=b(P_{\rm BD})$. Then letting $\eta_{\xi\text{-P},k}$ denote $\frac{\partial \xi_{\rm MB}}{\partial P_{\rm BD}}$ that is evaluated at iteration $k$, we have $b_k(P_{\rm BD})=\eta_{\xi\text{-P},k} P_{\rm BD}$, $B_k(P_{\rm BD})=\frac{1}{2} P_{\rm BD}^{\rm T} \eta_{\xi\text{-P},k} P_{\rm BD}$. Then, the new D-SP in (16) with a given $\lambda_{{\rm B},k}^{\rm sp}$ and $\eta_{\xi\text{-P},k}$ is simplified as

$$\min_{P_{\rm D},P_{\rm BD}} c_{\rm D}(P_{\rm D}) + (\lambda_{{\rm MB},k}^{\rm sp} - b_k(P_{{\rm BD},k-1}))^{\rm T} P_{\rm BD} + B_k(P_{\rm BD})$$
$$s.t. \; E_{P_{\rm D}} P_{\rm D} + E_{{\rm PB}_{\rm D}} P_{\rm BD} = d_{\rm D}, F_{P_{\rm D}} P_{\rm D} + F_{{\rm PB}_{\rm D}} P_{\rm BD} \geq e_{\rm D} \quad (24)$$

The $\eta_{\xi\text{-P},k}$ can be either solved from the sensitivity equations of the T-SP or evaluated via a fitting approach based on the preceding iteration data (cf. [29], and a comparison between these two approaches is also presented there). Typically, for this TDCED problem, this modified HGD algorithm converges faster and has a larger domain of convergence than the basic HGD algorithm does.

The tests in [29] demonstrate that in comparison with regular distributed algorithms like the ATC and the Biskas's method, the G-MSS typically requires about 8% or even less of the iterations that are needed by those algorithms. Furthermore, the TDCED will more accurately evaluate the LMPs of an ITD system as well as the relations between the transmission LMPs and the distribution LMPs. Hence, the TDCED and this distributed solution can be used for establishing an electricity market regarding an ITD system with high DER penetration.

## V. Conclusions

In this paper, a G-MSS method is suggested as a distributed solution to the TDCEM. Based on the heterogenous decomposition of the KKT conditions of the G-TDCM, two versions of HGD algorithms are presented. The basic HGD is simple and intuitive, while the modified HGD, though more complex in the formulation, typically converges faster, as it introduces a subsystem's response function to reduce the derivative of the composite mapping of the boundary variables. Both algorithms are proved to have sure optimality and convergence properties via the fixed-point theorem. The distributed G-MSS method is demonstrated to successfully solve a series of central functions, e.g. the PF, CA, VSA, ED and OPF, of the TDCEM, enabling distributed and effective cooperation between TSOs and DSOs to overcome the challenges arising from the DERs.

Future research directions include, for example, combing this G-MSS method with model predictive control to handle the uncertainties in an ITD system, and developing an effective and distributed heuristic approach to a mixed-integer G-TDCM from this G-MSS method.

## Appendix A
## Application of G-MSS to T-D SE [28]

A T-D SE (TDSE) model can be established as follows:

$$\min_{x_M, x_B, x_S} [Z_M - \hbar_M(x_M, x_B)]^T W_M [Z_M - \hbar_M(x_M, x_B)]$$
$$+ [Z_B - \hbar_B(x_M, x_B, x_S)]^T W_B [Z_B - \hbar_B(x_M, x_B, x_S)]$$
$$+ [Z_S - \hbar_S(x_B, x_S)]^T W_S [Z_S - \hbar_S(x_B, x_S)]$$
(25)

where the input $Z_M$, $Z_B$ and $Z_S$ are the measurements regarding master, boundary and slave subsystems, respectively; $\hbar_M$, $\hbar_B$ and $\hbar_S$ are the measurement functions associated with the subsystems, respectively, and the function $\hbar_B$ can be written as $\hbar_B = \hbar_{MB}(x_M, x_B) - \hbar_{BS}(x_B, x_S)$; $W_M$, $W_B$ and $W_S$ are weights.

To make the model in (25) consistent with the G-TDCM in (1), introduce a dummy variable $v_B$ and an additional constraint $v_B = Z_B - \hbar_B(x_M, x_B, x_S)$, and define

$$\begin{cases} z_M = [x_M; v_B], z_S = x_S \\ c_M(z_M, x_B) = [Z_M - \hbar_M(x_M, x_B)]^T W_M [Z_M - \hbar_M(x_M, x_B)] \\ \qquad + v_B^T W_B v_B \\ c_S(x_B, z_S) = [Z_S - \hbar_S(x_B, x_S)]^T W_S [Z_S - \hbar_S(x_B, x_S)] \\ f_B(z_M, x_B, z_S) = v_B - (Z_B - \hbar_B(x_M, x_B, x_S)) \end{cases}$$
(26)

Then, the model in (25) turns out to be a special case of the G-TDCM with $f_M = 0$, $f_S = 0$, $g_M = 0$, $g_S = 0$ and the $f_B$ being defined as below:

$$\begin{aligned} f_B &= v_B - (Z_B - \hbar_B(x_M, x_B, x_S)) \\ &= v_B - (Z_B - (\hbar_{MB}(x_M, x_B) + \hbar_{BS}(x_B, x_S))) \\ &= \underbrace{v_B - (Z_B - \hbar_{MB}(x_M, x_B))}_{f_{MB}(z_M, x_B)} - \underbrace{(-\hbar_{BS}(x_B, x_S))}_{f_{BS}(x_B, x_S)} \\ &= f_{MB}(z_M, x_B) - f_{BS}(x_B, x_S) \end{aligned}$$
(27)

Hence, the G-MSS method is applicable to the TDSE problem and a distributed solution can be derived in the way similar to what we have shown in Section IV. The detailed formulation of the subproblems are referred to the appendix of [28] and omitted here to save space.

## Appendix B
## Accuracy of Constructing a Response Function via Static Network Equivalencing

We will show that the $a_f$ constructed via (20) is accurate if the linear power flow equation in [31] is adopted in the G-TDCM and if a DPS is connected to the TPS via one boundary bus. To see this, let $\ell$ denote the set of the slave-subsystem buses connected to the boundary bus, and we have

$$s_{BS} = \sum_{j \in \ell} v_B \bar{Y}_{B,S,j} (\bar{v}_B - \bar{v}_{S,j}) \tag{28}$$

where $v_{S,j}$ is the complex voltage regarding the bus $j$ and $Y_{B,S,j}$ is the admittance of the branch from the boundary bus to the bus $j$. Further, define $Y_{B,S,j} = 0$ for any $j \notin \ell$, and we can transform (28) into (29) where $\mathbf{1}$ is the vector of all ones.

$$s_{BS} = v_B \bar{Y}_{B,S} (\mathbf{1}\bar{v}_B - \bar{v}_S) \tag{29}$$

Then, from the following linear power flow equation in [31]

$$v_S = v_B \left( \omega + \frac{1}{|v_B|^2} Y_{S,S}^{-1} \text{diag}\{\bar{\omega}\}^{-1} \bar{s}_S \right) \tag{30}$$

where $\omega = -Y_{S,S}^{-1} Y_{S,B}$, it follows that

$$\begin{aligned} s_{BS} &= |v_B|^2 \bar{Y}_{B,S} \left( \mathbf{1} - \bar{\omega} - \frac{1}{|v_B|^2} (\bar{Y}_{S,S})^{-1} \text{diag}\{\omega\}^{-1} s_S \right) \\ &= |v_B|^2 (Y_{B,S} \mathbf{1} + Y_{B,S} Y_{S,S}^{-1} Y_{S,B}) - \bar{Y}_{B,S} \bar{Y}_{S,S}^{-1} \text{diag}\{\omega\}^{-1} s_S \end{aligned}$$
(31)

In addition, we have

$$Y_{B,S}\mathbf{1} + Y_{B,S} Y_{S,S}^{-1} Y_{S,B} = \sum_{j \in \ell} Y_{B,S,j} + Y_{B,S} Y_{S,S}^{-1} Y_{S,B} = Y_{S,eq}. \tag{32}$$

Hence, it follows that

$$s_{\text{BS}} = |v_{\text{B}}|^2 \overline{Y_{\text{S,eq}}} - \overline{Y}_{\text{B,S}} \overline{Y}_{\text{S,S}}^{-1} \text{diag}\{\omega\}^{-1} s_{\text{S}}. \tag{33}$$

This means that the $a_f$ constructed via (20) is accurate for the linear power flow equation in [31], because in this case the total derivative of $f_{\text{BS}} = [\text{Re}(s_{\text{BS}}); \text{Im}(s_{\text{BS}})]$ with regard to $x_{\text{B}}$, which is derived from (33), equals $\frac{\partial a_f}{\partial x_{\text{B}}}$ that is computed based on the formula in (20).